\documentclass[showpacs,preprintnumbers]{revtex4}
\usepackage{amssymb}
\usepackage{amsmath}
\usepackage{graphicx}
\usepackage{dcolumn}
\usepackage{bm}
\usepackage{epsf}
\usepackage[dvips,usenames]{color}

\begin{document}

\preprint{HEP/123-qed}
\title{Three-neutron resonance trajectories for realistic interaction models}
\author{Rimantas Lazauskas}
\email{lazauskas@lpsc.in2p3.fr}
\affiliation{DPTA/Service de Physique Nucl\'eaire, CEA/DAM Ile de France, BP 12,
           F-91680 Bruy\`eres-le-Ch\^atel, France}
\author{Jaume Carbonell}
\email{carbonell@lpsc.in2p3.fr}
\affiliation{Laboratoire de Physique Subatomique et de Cosmologie, 53, avenue des
Martyrs, 38026 Grenoble Cedex, France.}
\homepage{http://isnwww.in2p3.fr/theo/Lazauskas/Eng/home.htm}
\homepage{http://isnwww.in2p3.fr/theo/Carbonell/Jaume.html}
\date{\today }
  %,21.60-n,27.10.+h

\begin{abstract}
Three-neutron resonances are investigated using realistic
nucleon-nucleon interaction models.
The resonance pole trajectories are explored
by first adding an additional interaction
to artificially bind the three-neutron system and then gradually removing it.
The pole positions for the three-neutron states up to $J$=5/2 are localized in the
third energy quadrant -- Im(E)$\leqslant0$, Re(E)$\leqslant0$ -- well before the additional
interaction is removed. Our study shows that realistic nucleon-nucleon
interaction models exclude 
any possible experimental signature of three-neutron resonances. 
\end{abstract}

\maketitle

\section{Introduction}

\label{sec:Intro}

The possible existence of pure neutron nuclei is a long standing
ambiguity in nuclear physics.
The neutron-neutron (\emph{nn}) scattering length is
negative and rather large $a_{nn}=-18.59\pm0.40$ fm~\cite{nn_scl4},
indicating that this system is almost bound in $^{1}\emph{S}_{0}$ state.
This value is actually the signature of a virtual bound state  just $\approx$100 keV
above the threshold. It is thus expected that by adding a few
neutrons, one can end up with a bound multineutron state, as it happens in
some fermionic systems, like  $^3$He atomic clusters~\cite{Guardiola}.
This is the reason for the recurrent turmoils appearing in the nuclear physics
society~\cite{Oglobin,Marques}.
However, the weakness of nuclear interaction in higher
partial waves (namely \emph{P} and \emph{D}), in comparison with their centrifugal
barriers, excludes the possibility of binding 'virtual' dineutrons together~\cite{Thesis}.
Recently, the non-existence of small
bound multineutron clusters has been settled out theoretically~\cite%
{Thesis,Pieper,Bertulani,Natasha}. Nevertheless, the existence of resonant
states of such nuclei having observable effects
had not yet been fully eliminated.

Indeed, in spite of numerous experimental and theoretical studies that
exploit different reactions and methods, the situation concerning
few-neutron resonances is not firmly established. One does not
have clear ideas even for the simplest case: the three-neutron
compound. A nice summary on the three-neutron system status up to
1987 can be found in~\cite{3N_analysis}. A few more recent
experimental studies have provided only controversial results.
In~\cite{He_3n_97}, the analysis of  $^{3}He(\pi^{-},\pi^{+})3n$ process
yielded no evidence of a three-neutron resonant state.
The explanation~\cite{3n_res_exp_70} of
double charge exchange differential cross sections
in $^{3}$He in term of a broad ($E=2-6i$ MeV)
three-neutron resonance were recently criticized by a more
thorough experimental study~\cite{3n_res_exp_86}.
Nevertheless the latter study further
suggested the existence of a wide $^{3}$n resonance at even larger energies
($E\approx20-20i$ MeV). Furthermore, very recent experimental results
on $^8He(d,{^6Li})4n$ reaction have shown some discrepancies with
what could be expected from phase space calculations, suggesting
the existence of a resonant tetraneutron~\cite{Exxon}.

There exists several theoretical efforts to find $^{3}$n and $^{4}$n
resonances. A variational study based on complex-scaling and simplified
nucleon-nucleon (\emph{NN}) interaction was carried out in~\cite{Csoto} with
the prediction of a $^{3}$n resonance at $E=14-13i$ MeV for the $J^{\pi}=3/2^{+}$ state.
On the other hand, no  $^{3}$n  nor  $^{4}$n
resonances were found by Sofianos et al.~\cite{Sofianos} using MT I-III
potential; only the existence of some broad subthreshold ($E_R<0$) states was
pointed out.
Realistic NN forces can provide however different
conclusions. These models contain interaction in \emph{P}- and higher
partial waves, which  -- due to the Pauli principle -- are a
crucial ingredient in binding fermionic systems.
The only study performed using realistic potentials was carried by Gl\"{o}ckle and
Witala~\cite{3n_res_gloe}. These authors were not able to find any
real three-neutron resonances. However due to some numerical instabilities,
the full treatment of $^{3}$n system was not accomplished and the
conclusions were drawn based on the phenomenological Gogny interaction~\cite{Gogny}.
Reference~\cite{Glockle_Hem} is probably the most complete
study of three neutron system. In this work, the trajectories for $^{3}$n states
with $J \leqslant $3/2 have been traced
by artificially enhancing a rank-2 separable nn interaction.
Our work is devoted to complement the studies of references
\cite{3n_res_gloe,Glockle_Hem}. We have explored
all $^{3}$n  states up to $J$=5/2,  fully relying on realistic \emph{NN} interactions.
Similarly to Ref.~\cite{Glockle_Hem}
we have used two different methods, namely complex-scaling (CS) and
analytical continuation in the coupling constant (ACCC), to trace the resonance
trajectories. This allows us to check the reliability of our results, as well as to
judge on the pertinence of the two methods used.

{\it [Calculations presented in this paper use $\frac{\hbar ^{2}}{m}=41.44$ MeV$\cdot $fm$^{2}$ as an imput for the mass of the neutrons.]}

\section{Theoretical background}\label{sec:Theorie}

Resonance eigenfunctions correspond to complex energy solutions of the
Schr\"{o}dinger equation:%
\begin{equation}
\widehat{H}\Psi_{res}=E_{res}\Psi_{res},\qquad E_{res}=\varepsilon_{res}%
-\frac{i}{2}\Gamma_{res}.%
\end{equation}

Since physical resonance have positive energy real parts, $\varepsilon_{res}>0$,
the corresponding eigenfunctions are not square integrable.
Nevertheless, by applying them a similarity transformation
they can by mapped onto normalizable states. That is:
\begin{equation}
\left(  \widehat{S}\widehat{H}\widehat{S}^{-1}\right)  \left(  \widehat{S}%
\Psi_{res}\right)  =E_{res}\left(  \widehat{S}\Psi_{res}\right)
\end{equation}
with $\widehat{S}\Psi_{res}\rightarrow0$ as $r\rightarrow\infty$.

Functions $\left(\widehat{S}\Psi_{res}\right)$ are in Hilbert space,
although $\Psi_{res}$ are not.
The complex-scaling method  is defined by means of the similarity
operator~\cite{Moiseyev,Ho}:%
\begin{equation}
\widehat{S}=e^{i\theta r\frac{\partial}{\partial r}},%
\end{equation}
such that any analytical function $f(r)$ is transformed according to:
\begin{equation}
\widehat{S}f(r)=f(re^{i\theta}). \label{Comp_sc_oper}
\end{equation}

For a broad class of potentials, complex-scaling operation does not affect the bound
and resonant state spectra of the Hamiltonian $\widehat{H}$, provided
$0\leqslant\theta<\frac{\pi}{2}$. However the continuous spectra of $\widehat{H}$ will
be rotated by an angle $2\theta$.
Resonance eigenfunctions $\left(\widehat{S}\Psi_{res}\right) $
of the scaled Hamiltonian become square integrable
if \ $2\theta>\left\vert \arg(E_{res})\right\vert$ and therefore the standard
bound state techniques can be applied to determine the corresponding eigenvalues.

\bigskip
In order to solve the three-body problem we use Faddeev equations
in configuration space \cite{Fadd_art}, first derived by Noyes~\cite{Fadd_art,Noyes}.
Though this formalism was initially developed to investigate
the three-body continuum,
it turns to be very useful to treat bound-state problems as well.
For  three identical particles,  the Faddeev-Noyes equations read:
\begin{equation} \label{Fad_eq}
\left(  E-\widehat{H}_{0}-V_{ij}\right)  \psi_{ij,k}=V_{ij}(P^{+}+P^{-})\psi_{ij,k}.
\end{equation}
$\widehat{H}_{0}$ is the three-particle kinetic energy operator, $V_{ij}$
the two-body force, $\psi_{ij,k}$ the Faddeev component and $P^{+},P^{-}$
denotes cyclic particle permutation operators.
The properly symmetrized three-body wave function is
$\Psi=(1+P^{+}+P^{-})\psi_{ij,k}$.
In order to simplify the kinetic energy operator and to separate internal
and center of mass degrees of freedom we use  Jacobi
coordinates: $\overrightarrow{x_{ij}}=\overrightarrow{r_{j}}-\overrightarrow
{r_{i}}$ and $\overrightarrow{y_{ij}}=\frac{2}{\sqrt{3}}\left[
\overrightarrow{r_{k}}-\frac{1}{2}\left(  \overrightarrow{r_{i}}%
+\overrightarrow{r_{j}}\right)  \right]  $.

Complex-scaling  Faddeev equations causes no difficulties: one has simply  to
scale all the Jacobi vectors with the same exponential factor:
\begin{equation}
\overrightarrow{x_{ij}}\rightarrow\overrightarrow{x_{ij}}e^{i\theta}\qquad \text{and\qquad}
\overrightarrow{y_{ij}}\rightarrow\overrightarrow{y_{ij}}e^{i\theta}.%
\end{equation}

Such transformation affects only the hyperradius $\rho=\sqrt{x_{ij}^{2}+y_{ij}^{2}}$,
whereas neither the expressions of permutation operators -- $P^{+},P^{-}$ -- nor
the angular dependence of Faddeev equations are affected.
Using Jacobi coordinates,
the kinetic energy operator can be expressed as a six-dimensional Laplacian
$\widehat{H}_{0}=-\frac{\hbar^{2}}{m}\Delta_{\chi}$ with $\chi\equiv
(\overrightarrow{x_{ij}},\overrightarrow{y_{ij}})$ and equation (\ref{Fad_eq})
transforms into:
\begin{equation}
\left[  E+e^{-i2\theta}\Delta_{\chi}-V_{ij}\left(  x_{ij}e^{i\theta}\right)
\right]  \widetilde{\psi}_{ij,k}(\overrightarrow{x_{ij}},\overrightarrow
{y_{ij}})=V_{ij}\left(  x_{ij}e^{i\theta}\right)  (P^{+}+P^{-})\widetilde
{\psi}_{ij,k}(\overrightarrow{x_{ij}},\overrightarrow{y_{ij}})
\label{SC_Fad_eq}%
\end{equation}

As in our previous works~\cite{Thesis,Jaume_Fred_PRC}, equation (\ref{SC_Fad_eq}) results
into a system of integrodifferential equations by first expanding
and then projecting the Faddeev components into a -- partial wave -- basis of spin,
isospin and angular momentum:
\begin{equation}\label{PWB}
\widetilde{\psi}_{ij,k}(\overrightarrow{x_{ij}},\overrightarrow{y_{ij}})=%
{\displaystyle\sum\limits_{\substack{STL\\l_{x}l_{y}}}}
\frac{\psi_{k,l_{x}l_{y}}^{STL}(x_{ij},y_{ij})}{x_{ij}y_{ij}}\left[
e^{ST}\otimes\mathcal{Y}_{l_{x}l_{y}}(\widehat{x_{ij}},\widehat{y_{ij}%
})\right]  _{JT}.%
\end{equation}

Next, the radial dependence of  amplitudes $\psi_{k,l_{x}%
l_{y}}^{STL}(x_{ij},y_{ij})$ is  developed  in a basis of cubic
Hermite splines~\cite{Boor_book}. Such a procedure applied to the
complex scaled Faddeev equation (\ref{SC_Fad_eq}) results into a
generalized eigenvalue algebraic problem:
\begin{equation}
A \; X=E_{res}\; B \; \label{Lin_alg_eq}%
\end{equation}
where $A$ and $B$ are known complex matrices, $E_{res}$ and $X$
are respectively the complex eigenvalue and eigenvector to be determined.

The partial wave expansion (\ref{PWB}) is pushed up to amplitudes with intermediate angular
momenta $max(j_x,j_y)<$4.5, which guarantees at least three-digit accuracy
in the results. This requires solving linear system (\ref{Lin_alg_eq})
of a relatively large ($n_{eq}\sim10^{5}-10^{6}$)
size, which prevent us applying direct linear algebra methods.
In order to avoid cumbersome matrix inversion, we use
inverse-iteration techniques to search only individual
eigenvalues  and iterative methods to solve the linear systems.
Technical details of the numerical methods in use can be found in~\cite{Thesis}.

\bigskip
There is an apparently simpler procedure to depict resonance
trajectories, namely the method of analytic continuation in the coupling
constant (ACCC). This method, developed by V.I. Kukulin et al.~\cite%
{Kukulin_JPA77,Kukulin_Sov79_29,Kukulin_PL78,Kukulin_Sov79_30,Kukulin_book},
is based on the intuitive argument that a resonance may be treated
as an eigenstate  which arises from a bound state when
the intensity of the attractive part of the interaction decreases.
The S-matrix pole of a resonant state is defined as the analytic
continuation of a bound-state pole in the
coupling constant of the attractive part of the Hamiltonian.
One assumes that the Hamiltonian can be written as
$H(\lambda)=H_{0}+\lambda  H_{att}$, where $H_{att}$ is the
attractive part of the unperturbed Hamiltonian $H(\lambda=1)$. If for
some value of $\lambda$ the Hamiltonian has a bound state, then by
gradually decreasing $\lambda$, its binding energy decreases and the state reaches
the threshold at $\lambda=\lambda_{0}$ -- i.e. $E(\lambda=\lambda _{0})=0$ --
to become a resonant or a virtual one.

For physical Hamiltonians, the  binding energy is assumed to be analytical in $\lambda$.
Moreover, it can be shown~\cite{Kukulin_book},
that for a two-body system the square root of the binding
energy -- $k_{\ell}=\sqrt{-E}$ -- behaves near the threshold $\lambda=\lambda _{0}$ as:

\begin{equation}
k_{\ell}\sim \left\{
\begin{array}{c}
\lambda -\lambda _{0} \\
\sqrt{\lambda -\lambda _{0}}%
\end{array}%
\right.
\begin{array}{c}
\text{for\quad }\ell=0 \\
\text{for\quad }\ell\neq 0%
\end{array}
\label{ACCC_ff}
\end{equation}

By introducing the complex variable
\[ x \equiv \left\{
\begin{array}{c}
\lambda -\lambda _{0} \\
\sqrt{\lambda -\lambda _{0}}%
\end{array}%
\right.
\begin{array}{c}
\text{for\quad }\ell=0 \\
\text{for\quad }\ell\neq 0%
\end{array}  \]
one can consider $k_{\ell}$ as a function of variable $x$ and
analytically continue it from the bound state ($\lambda
>\lambda _{0}$) to the resonance region ($\lambda <\lambda _{0}$). Motivated
by the functional form (\ref{ACCC_ff})  near the threshold,
we use for $k_{\ell}$ the Pad\'{e} approximant~\cite{Baker_Pade}:

\begin{equation}
k_{l}(x)=k^{\left[ N,M\right] }(x)=\frac{a_{1}\cdot x+a_{2}\cdot
x^{2}+\ldots a_{N}\cdot x^{N}}{1+b_{1}\cdot x+b_{2}\cdot x^{2}+\ldots
b_{M}\cdot x^{M-1}}  \label{Pade}
\end{equation}

In general, for  systems with $n>2$, $k_{\ell}$ is the
relative momenta to the nearest desintegration threshold
$k_{\ell}=\sqrt{-(E_{n}-E_{i<n})}$. Contrary to the n=2 case, for $n>2$
systems the angular momenta does not determine anymore if a bound state
turns into a virtual or resonant one when $\lambda <\lambda _{0}$.
Nevertheless, according to (\ref{ACCC_ff}) this transition can
be discriminated by studying the nearthreshold behavior of $\sqrt{-(E_{n}-E_{i<n})}$.
If $(E_{n}-E_{i<n})$ is linear in $(\lambda -\lambda _{0})$ at
the origin, the bound state turns into a resonance; virtual state appears if $%
(E_{n}-E_{i<n})$ is quadratic in $(\lambda -\lambda _{0})$.

\section{Three-neutron resonances}\label{sec:Results}

Complex scaling methods (CS) turns out to be a very powerful tool when treating
resonance problems in atomic and molecular systems. However scaled nuclear
potentials introduce numerical instabilities not encountered in atomic
physics,  which is dominated by Coulomb (or Coulomb derived) potentials.
The long range
part of the realistic nuclear potentials decreases exponentially with
a Yukawa pion-tail. Using CS, exponentially decreasing function $e^{-\mu r}$
transforms into $e^{-\mu\lbrack cos\theta+isin\theta]r}$.
For a scaling parameter $\theta>0$, the potential range increases as $1/cos\theta$, introducing
sizeable oscillations and demanding larger and denser grids to describe the system.
These difficulties can be partially avoided by extending $\theta$ to complex values,
as proposed by Gl\"{o}ckle et al.~\cite{3n_res_gloe}.

\begin{figure}[h!]
\begin{center}
\includegraphics[width=14.cm]{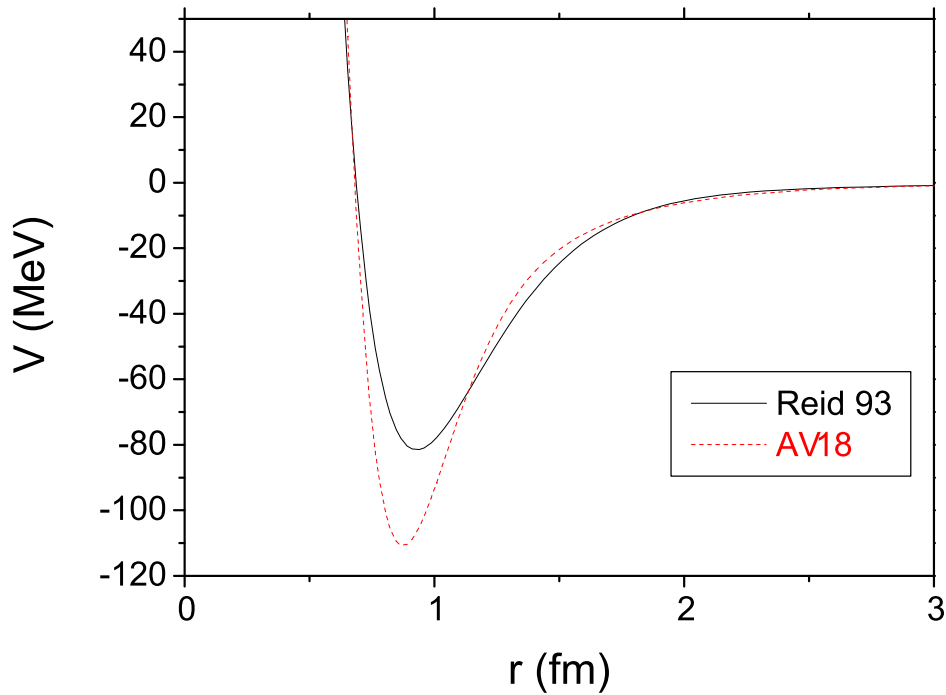} \vspace{-1.2 cm}
\end{center}
\caption{Reid 93 and AV18 $^{1}\emph{S}_{0}$ \emph{nn} potentials.}
\label{Fig:Reid_AV18}
\end{figure}

\begin{figure}[h!]
\begin{center}
\epsfxsize=8.4cm\mbox{\epsffile{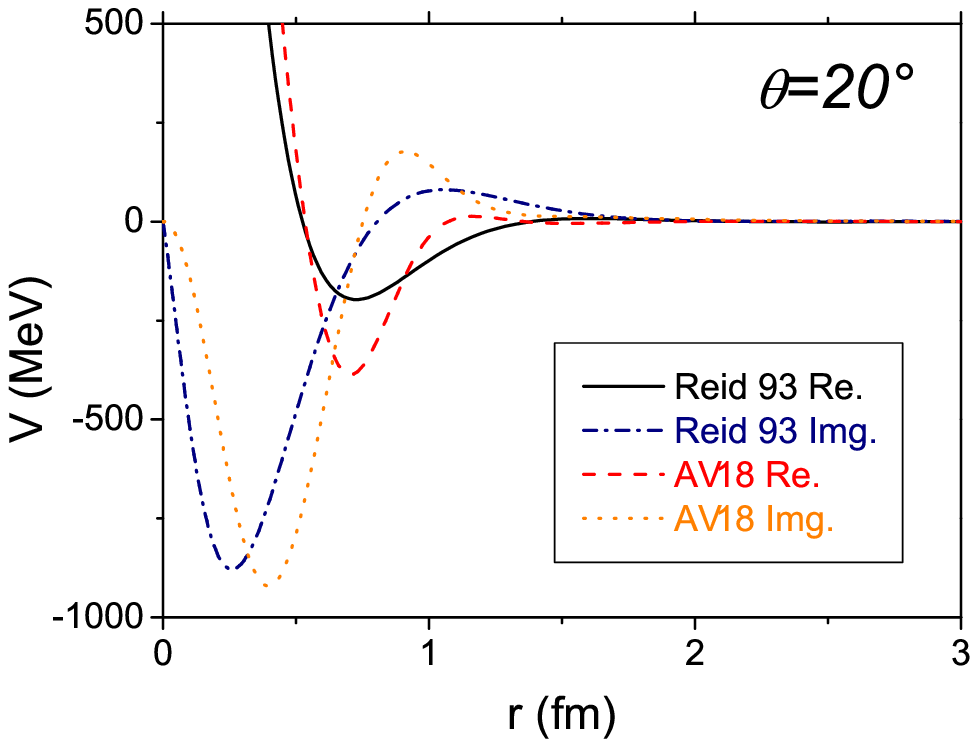}}\hspace{-0.6cm} %
\epsfxsize=8.4cm\mbox{\epsffile{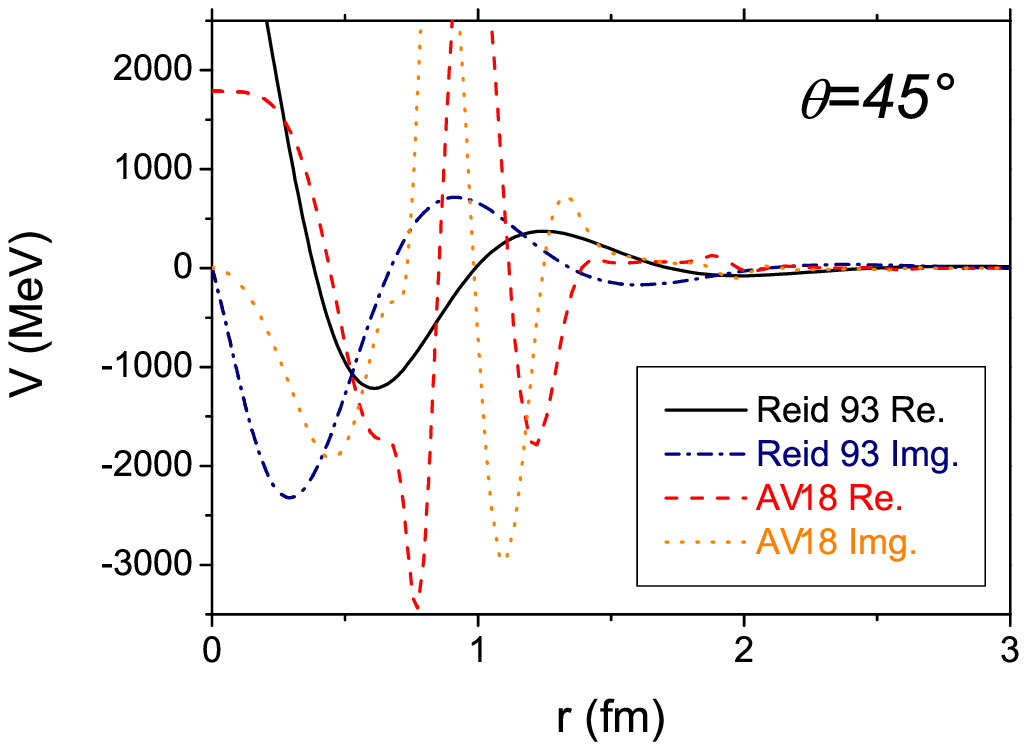}}
\end{center}
\vspace{-1.2 cm}
\caption{Analytical continuation of the $^{1}\emph{S}_{0}$ Reid 93 and AV18
\emph{nn} potential under the complex scaling transformation eq. (\protect\ref%
{Comp_sc_oper}) with $\protect\theta=20^\circ$ and $\protect\theta%
=45^\circ$. One can notice that AV18 potential results into much
more oscillating structure than Reid 93.}
\label{Fig:Comp_rot_pot}
\end{figure}

A much more serious problem arises due to the hard-core
short range part of the nuclear potentials (see Figs.~\ref{Fig:Reid_AV18}).
Notably, the presence of the $e^{{%
-cr^{2}}}$ terms in AV18~\cite{AV18} and AV14 \cite{AV14} models
settles that for CS with $\theta>45^\circ$ these potentials become divergent.
Furthermore even for smaller $\theta$ these two models
have a strong oscillatory behavior
which makes numerics unstable.
This can be seen in Fig.~\ref{Fig:Comp_rot_pot} where
the transformed NN potentials for two different angles are plotted.
Note that  the number
of oscillations as well as their amplitudes increase dramatically with the rotation angle.
In this sense Reid 93 model~\cite{Reid} has the best analytical properties with respect to CS,
though it provides stable results only for the $\theta\lesssim35^\circ $.

When treating two-body systems, the numerical complications described above can
be avoided by using the so-called `smooth-exterior' complex scaling (SECS)~\cite{Moiseyev}.
The problem due to the hard core of the nuclear potentials is
overcome by performing scaling operation only in the outside region of the
interaction. This method can be implemented by using the following
transformation~\cite{3n_res_gloe,Moiseyev}:%
\[ r\rightarrow F(r) \]
with
\begin{equation}
F(r)=r+\left[  e^{i\theta}-1\right]  \left\{  r+\frac{1}{4\eta}\ln
\frac{\left[  1+e^{2\eta(r-R)}\right]  \left[  1+e^{-2\eta(r-R)}\right]
}{\left[  1+e^{2\eta(r+R)}\right]\left[  1+e^{-2\eta(r+R)}\right]  }\right\}  .
\end{equation}
where $\eta$ and $R$ are smoothing parameters.
However, despite the efficiency of this method when dealing with n=2 problem, it is
not easy to implement it in the n=3 Faddeev equations~\cite{3n_res_gloe}.
The problems turned out to be crucial when calculating deeply lying
resonances and we were limited to those with $E_{re}/E_{img} \succsim0.5$.

\bigskip

All the difficulties discussed above are not encountered in the ACCC method,
which can be used in principle even to determine subthreshold ($\varepsilon_{res}<0$)
resonances. In calculating the input binding energies for
determining the Pad\'e approximant
(\ref{Pade}), one can also scale the total \emph{nn} potential with a scaling factor
$\gamma$ -- not necessarily only its
attractive part -- and consider $\gamma$ as the extrapolation parameter $\gamma\equiv\lambda$.

\vspace{-.8cm}
\begin{figure}[h!]
\begin{center}
\includegraphics[width=14.cm]{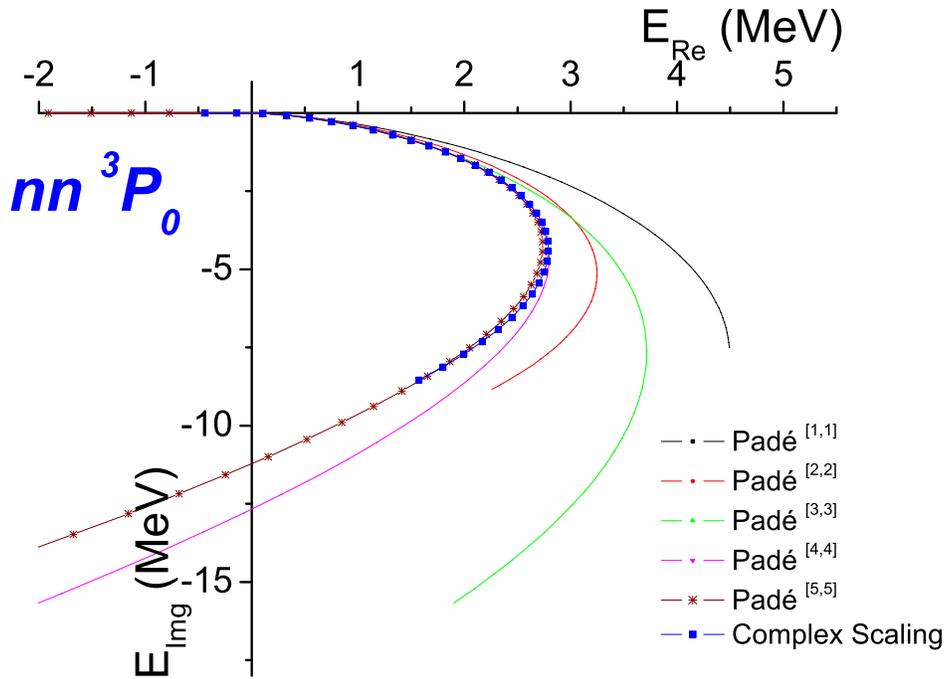} \vspace{-1.2 cm}
\end{center}
\caption{Comparison between ACCC and CS methods for $^{3}\emph{P}_{0}$
\emph{nn} resonance trajectories with Reid93 potential.
ACCC results with several Pad\'e orders [N,M] and $\gamma$=6.1  to 1.0 are represented by solid lines.
CS results are denoted by squares
and correspond to  $\gamma$ from 6.1 to 2.7 by steps of 0.1.
Star points correspond to [5,5] Pad\'e approximant used in ACCC.
They are already very close to CS results:  by adding a few terms in the approximant
a perfect agreement is reached.}
\label{Fig:2n_ACCC_CS}
\end{figure}

In Figure \ref{Fig:2n_ACCC_CS} we compare the  results of various order Pad\'{e}
approximant used in (\ref{Pade}) with the ones obtained using CS method: dineutron
resonance trajectory is depicted for Reid 93 $^{3}\emph{P}_{0}$ \emph{nn}-waves.
One can see that a nice agreement is reached between both techniques even
when calculating broad resonances,
near and beyond the saturation point (where real energy part
starts decreasing), provided high order Pad\'{e} approximant are used.
This requirement can not always be  met in
numerical calculations with realistic interactions, since it
implies a very accurate binding energy input.
Another critical point of ACCC method, as discussed in~\cite{Kukulin_book},
is that its efficiency highly depends on the accuracy of $\lambda_0$ value.
The determination of $\lambda_0$ is often difficult, since one has to deal with
critically bound  systems  having very extended wave functions.
Therefore one has to be rather prudent when applying this method and always check if
Pad\'{e} extrapolation converges.

\begin{table}[h!]
\caption{Critical enhancement factors $\protect\gamma_c$ required to bind
dineutron in various states and for different \emph{NN} realistic interaction
models in use.}
\label{gamma_2N}%
\begin{tabular}{l|llll}
\hline
& Nijm II & Reid 93 & AV14 & AV18 \\ \hline\hline
$^{2}n(^{1}S_{0})$ & 1.088 & 1.087 & 1.063 & 1.080 \\
$^{2}n(^{3}P_{0})$ & 5.95 & 5.95 & 5.46 & 6.10 \\
$^{2}n(^{3}PF_{2})$ & 3.89 & 4.00 & 4.30 & 4.39 \\
$^{2}n(^{1}D_{2})$ & 9.28 & 9.22 & 9.54 & 10.20 \\ \hline
\end{tabular}%
\end{table}

Before starting to analyze three-neutron system it could be useful to discuss the
basic properties of dineutron and \emph{nn} interaction in general. As mentioned
above, dineutron is almost bound in the $^{1}\emph{S}_{0}$ state: one should
enhance the nuclear potential only by a factor $\gamma_c\approx1.08$ to make it
bound (see Table~\ref{gamma_2N}).
Notice the very good agreement for the critical
enhancement factors $\gamma_c$ obtained by different local \emph{NN}-interaction
models. Only AV14  slightly deviates from the other models, due
to its charge invariance assumption: the potential being adjusted to
reproduce neutron-proton (\emph{np}) scattering data, ignores the fact that experimental $^{1}\emph{%
S}_{0}$ \emph{nn} scattering length is smaller in magnitude than \emph{np} one~\cite{nn_scl4}.
The spherical symmetry of this state
makes that when $\gamma\to1$, i.e. for the physical value of the
potential, the bound state pole moves staying on the imaginary
$k$-axis and becomes a virtual state, not a resonance. The
approximate position of this state can be evaluated from the \emph{nn}
scattering length by means of the relation $E_{virt}\approx\frac{\hbar^2}{ma_{nn}^2}$.
The results thus obtained together with the exactly calculated virtual state energies
are given in Table~\ref{long_de_diff}.
\begin{table}[h]
\caption{Nuclear model predictions for \emph{nn} scattering length (in fm) as well as corresponding virtual state (in MeV), evaluated
from scattering length and calculated exactly.}
\label{long_de_diff}%
\begin{tabular}{l|llll}\hline
& Nijm II & Reid 93 & AV14 & AV18 \\ \hline\hline
$a_{nn}(^{1}S_{0})$ & -17.57 & -17.55 & -24.02 & -18.50 \\
${\hbar^2}/(ma^{2}_{nn})$ & 0.134 & 0.135 & 0.072 & 0.121 \\
$E_{virt}(^{1}S_{0})$ & 0.1162 & 0.1165 & 0.0647 & 0.1055 \\ \hline
\end{tabular}%
\end{table}

In fact, multineutron physics being in low energy regime, is
dominated by the large \emph{nn} scattering length value ($a_{nn}$).
The wave function has only a small part in the interaction region
(effective range $r_0<<a_{nn}$) and therefore marginally depends on the particular
form that the \emph{nn} $^{1}\emph{S}_{0}$-potential  can take, once
$r_0$ and $a_{nn}$ are fixed~\cite{Thesis}. On the other hand
$r_0$ is controlled by the theory -- the pion Compton-wavelength  -- and
$a_{nn}$ is constrained by the experiment. These effective range
theory arguments~\cite{Platter} show that one should not rely on
the modifications of $^{1}\emph{S}_{0}$ waves  to favor
the possible existence of bound or resonant multineutron states.

\emph{P}-waves of \emph{nn} interaction are extremely weak and this turns to be a
major reason why multineutrons are not bound~\cite{Thesis}. Neutron-neutron
interaction in $^{3}\emph{P}_{1}$ channel is even repulsive, whereas
potentials in $^{3}\emph{PF}_{2}$ and $^{3}\emph{P}_{0}$
channels should be multiplied by considerably large factors -- $\gamma=[3.9-4.4]
$ and $[5.5-6.1]$ respectively (see Table~\ref{gamma_2N}) -- to force
dineutron's binding. The centrifugal barrier moves these artificially
bound states into resonances when  $\gamma$ is slightly reduced from its critical value.

\begin{figure}[h!]
\begin{center}
\includegraphics[width=14.cm]{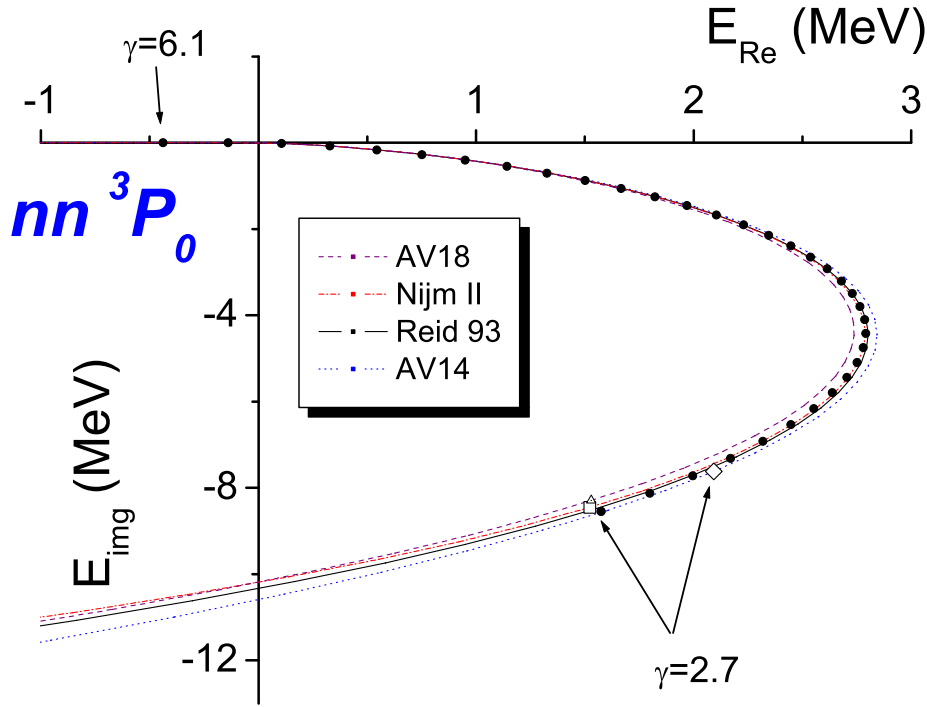} \vspace{-1.2 cm}
\end{center}
\caption{Dineutron $^{3}\emph{P}_{0}$ resonant trajectories in complex
energy plane for  AV14, NijmII, Reid 93 and AV18 \emph{nn} interactions.
Solid dot symbols correspond to $\protect\gamma$ values from 6.1 to 2.7 by steps of 0.1 obtained
with CS method and Reid 93. Continuous lines represent ACCC results.
ACCC and CS results superimpose up to $\gamma$=2.7 point, the limit of CS  applicability.}
\label{Fig:2n_res_3P0}
\end{figure}

\begin{figure}[h!]
\begin{center}
\includegraphics[width=14.cm]{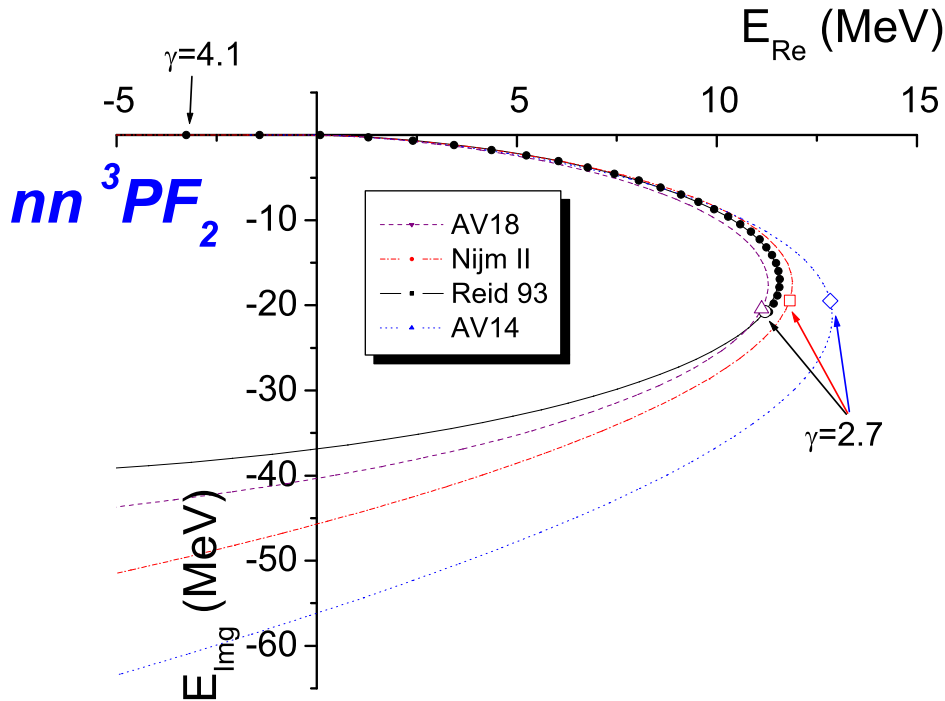} \vspace{-1.2 cm}
\end{center}
\caption{Dineutron $^{3}\emph{PF}_{2}$ resonant trajectories in complex energy plane, displayed using
the same conventions as in figure \ref{Fig:2n_res_3P0}.}
\label{Fig:2n_res_3PF2}
\end{figure}

\bigskip
Dineutron resonance trajectories for  $^3P_0$ and $^3PF_2$ states  are displayed
respectively in Figures \ref{Fig:2n_res_3P0} and \ref{Fig:2n_res_3PF2}.
They have been obtained with  CS and ACCC methods and different NN interaction models.
By using a high order Pad\'e approximant and accurate
${^2n}$ binding energies input in the ACCC method, we observe a perfect agreement between these
two techniques. Due to the limitations in the rotation angle $\theta$,
CS method was applied only up to $\gamma=2.7$. The corresponding resonance positions obtained with different 
 NN models are explicitly indicated in the figures.
One should note that the resonance trajectories in both states have  similar shapes.
When decreasing $\gamma$ from its critical value $\gamma_c$,
the resonance width starts departing slowly from zero and then increases linearly.
On the other hand,
the real part of the resonance energy -- $\varepsilon_{res}$ -- first grows linearly with
$\gamma_c-\gamma$. Afterwards, this growing saturates and reach a maximum
value from where it quickly decreases, vanishes and becomes negative.
ACCC method allows to follow the resonance trajectories up to the physical
value $\gamma=1$ at which the dineutron states are deeply subthreshold.
However, the transition to the third energy quadrant happens at a value $\gamma'_c$
well above.
Some resonance trajectory properties obtained using ACCC method
are summarized in Table~\ref{gamma_res_2N}. 
Note that in none of  Figures \ref{Fig:2n_res_3P0} and \ref{Fig:2n_res_3PF2}
the resonance trajectories are plotted until the physical value $\gamma=1$.

It is clear from results given in Table~\ref{gamma_res_2N} that
the existence of any observable $\emph{P}$-wave dineutron resonance is excluded:
only subthreshold poles with large widths persist, making such structures physically meaningless.
It is worth noticing recent few-nucleon scattering calculations indicating that a good description of $3N$ and $4N$
scattering observables, would require stronger \emph{NN} $\emph{P}$-waves\cite{P_waves_Gloe,P_waves_Pisa,Thesis}.
However the  modifications involved -- less than 20\% -- let the dineutron resonances still subthreshold

\begin{table}[h]
\caption{Enhancement factor values ($\gamma'_c$) at which dineutron resonances become subthreshold
($\varepsilon_{res}=0$), and imaginary energy values E$_{img}(\gamma'_c)$ at this point (MeV).
Resonance energy E$_{res}$ for physical \emph{nn} interaction
(i.e. at $\gamma=1$) obtained using ACCC method.}
\label{gamma_res_2N}%
\begin{tabular}{l|cccc|cccc}\hline
    & \multicolumn{4}{c|}{$^{3}P_{0}$} & \multicolumn{4}{c}{$^{3}PF_{2}$} \\  \cline{2-9}
                     & Nijm II & Reid 93 & AV14 & AV18  & Nijm II & Reid 93 & AV14    & AV18 \\ \hline\hline
$\gamma'_c$          & 2.27    & 2.26   & 2.08 & 2.24  & 1.64    &  1.71   &  1.46   & 1.73  \\
E$_{img}(\gamma'_c)$ & -10.2   & -10.3  & -10.6& -10.2 & -45.6    & -36.9   & -56.2   & -40.3   \\
E$_{res}(\gamma=1)$&-14.1-17.2\textit{i}&-14.2-18.5\textit{i}&-10.3-18.1\textit{i}&-12.1-18.0\textit{i}
           & -20.5-64.8\textit{i}& -15.9-39.9\textit{i}& -17.9-80.1\textit{i}&-34.1-45.4\textit{i}\\
\hline
\end{tabular}%
\end{table}

One should remark the astonishing similarity for the
\emph{P}-wave dineutron resonance trajectories obtained with
\emph{NN} potientials having quite a different shapes (see Fig.~\ref{Fig:Reid_AV18}).
$^{3}\emph{P}_{0}$ curves for the three charge symmetry-breaking models considered superimpose,
whereas for  $^{3}\emph{PF}_{2}$ they separate only when very large resonance energies are reached.
Enhancement factors employed in tracing these curves are unphysically large and
produce very broad resonances: $^{3}\emph{P}_{0}$ state moves into third energy quadrant at
$E_{img}\sim10$ MeV, while in $^{3}\emph{PF}_{2}$ case this value goes beyond 35 MeV.

\bigskip
Two neutron states with orbital angular momentum $\ell$=2, can be realized only in singlet state ($^{\emph{1}}$\emph{%
D}$_{\emph{2}}$). This state
is dominated by a sizeable centrifugal barrier and the critical enhancement factors to bind dineutron
($\gamma_c$) is considerably large (see Table~\ref{gamma_2N}).
The central effective potentials
\begin{equation*}
V_{eff}(r)=V_{nn}(r)+\frac{\hbar^{2}}{m_{n}}\frac{\ell(\ell+1)}{r^{2}}
\end{equation*}
in this and higher angular momentum \emph{nn} partial waves are smoothly decreasing functions,
without any dips, implying without further calculation that dineutron can not
have $\ell\geqslant$2 observable resonances in the 4th quadrant.

\bigskip
As mentioned in section~\ref{sec:Theorie} we cannot obtain numerically
all the eigenvalue spectra of the 3n problem. Only a few specific
eigenvalues of the linear system (\ref{Lin_alg_eq}) can be extracted by
applying iterative methods.
When using the CS method, these techniques do
not allow us a-priori to separate eigenvalues related to the resonances from
the spurious ones related to the rotated continuum.
In order to force the numerical procees to converge towards the resonance position
we have to initialize it with a rather accurate guess value.
This obliged us to follow the
procedure described in~\cite{3n_res_gloe}: first we bind three neutrons by
artificially making \emph{nn} interaction stronger, and then we gradually remove
the additional interaction and follow the trajectory of this state. Note, that
in bound state calculations one can use linear algebra methods
determining the extreme eigenvalues of the spectra (e.g. Lanczos or Power-method),
whereas resonance eigenvalue in the CS matrices is not anymore an extreme one.

By enhancing the $^{\emph{1}}\emph{S}_{\emph{0}}$ \emph{nn}-potential
there is  no way to bind three neutrons without first binding dineutron.
On the other hand, as quoted before, this wave is controlled both by
theory and experiment and modifications of its form can not affect multineutron physics.
Three-neutron can neither be bound if we
keep the $^{\emph{1}}$\emph{S}$_{\emph{0}}$ interaction unchanged,
and multiply all \emph{nn} \emph{P}-waves with the same enhancement
factor; by doing so, dineutron will be bound in $^{3}\emph{PF}_{2}$ channel before three-neutron was.
In view of that,  we have tried to enhance only one of $\emph{P}$ interactions, keeping the
usual strengths for the other ones.
The $^{3}\emph{P}_{1}$ potential is purely repulsive and its enhancement can not give any positive effect.
The $^{3}\emph{P}_{0}$ enhancement  gives null result as well: dineutron is always
bound before any  $^{3}n$ states is formed.
Only enhancing $^{3}\emph{PF}_{2}$ channel we managed to
bind trineutron  without first binding dineutron and this happens in the $J^{\pi}=\frac{3}{2}^{-}$ $^{3}n$
state alone.
These properties turned to be general in the four
realistic interactions (AV14, Reid 93, Nijm II and AV18) we have investigated.

The critical $^{3}\emph{PF}_{2}$ enhancement factors $\gamma_c$ required to bind ${^3n}$ are summarized in
Table~\ref{gamma_3n}. They are so large that dineutron, although unbound,
is already resonant; the
corresponding resonance positions are summarized in the bottom line of this Table.
One can remark, once again, the rather good agreement between the
different model predictions. This fact, as well as the similarity in the dineutron case,
predictions suggest that the different realistic local-interaction models would
provide a good qualitative agreement in multineutron physics as well.
Therefore, in further analysis of three-neutron resonances we will rely on a single interaction model.
In this respect,  Reid 93 model is the most suited, since it possess the best
analytical properties and consequently provides the most stable numerical results for CS method.
\begin{table}[h!]
\caption{Critical enhancement factors $\protect\gamma_c$ required for $^{3}%
\emph{PF}_{2}$ \emph{nn} channel to bind $J^{\protect\pi}=\frac{3}{2}^{-}$
three-neutron and  corresponding $J^{\protect\pi}=2^{-}$ dineutron resonance positions (MeV).}\label{gamma_3n}%
\begin{tabular}{l|llll}\hline
                   & Nijm II & Reid 93 & AV14 & AV18 \\ \hline\hline
$\gamma_c({^{3}n}) $ & 3.61 & 3.74 & 3.86 & 3.98 \\
E($^{2}n$) MeV     & 5.31-2.41$i$ & 5.41-2.52$i$ & 5.20-2.49$i$ &4.83-2.31$i$ \\ \hline
\end{tabular}
\end{table}

\begin{figure}[h!]
\begin{center}
\includegraphics[width=14.cm]{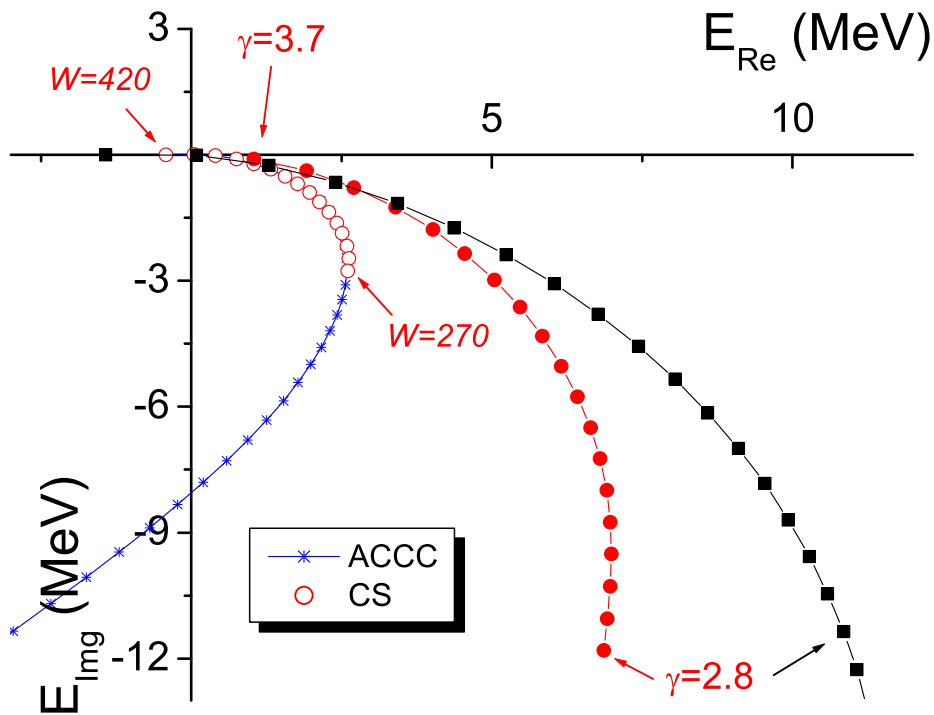} \vspace{-1.2 cm}
\end{center}
\caption{$J^\protect\pi=3/2^-$ three-neutron state resonance trajectory
obtained when reducing the strength $W$ of phenomenological Yukawa-type force
(open circles for CS and solid line+star points for ACCC methods). Trajectory
depicted by full circles represents one obtained using CS, when reducing enhancement factor $\gamma$ for $%
^{3}\emph{P}_{2}-^{3}\emph{F}_{2}$ \emph{nn} interaction .
Trajectory
depicted by full squares is dineutron resonance path in $^{3}\emph{P}%
_{2}-^{3}\emph{F}_{2}$ channel, obtained by enhancing \emph{nn}-interaction in
these waves. Presented results are based on Reid 93 model.}
\label{Fig:3_2_nnn_traj}
\end{figure}

As mentioned, above only a $\frac{3}{2}^{-}$ three-neutron state can be
bound by enhancing single \emph{NN} interaction, without first binding dineutron.
In Figure~\ref{Fig:3_2_nnn_traj}  we trace  the  $^3n$ resonance trajectory 
(full circles) for this state when the  $^{3}\emph{PF}_{2}$ enhancement factor 
changes from 3.7 to 2.8 with step of 0.05.
Result were obtained using CS methods. 
Extending the calculations to smaller $\gamma$ values generated
numerical instabilities,  due to the necessity to scale
Faddeev equations with ever increasing $\theta$.  
It can be seen however, that this trajectory bends faster than the analogous one
for the dineutron in $^{3}\emph{PF}_{2}$ state, therefore
indicating that it will finish in the third energy quadrant with Re(E)$<$0.

Three-neutron can also be bound in states $\frac{3}{2}^{+}$ and $\frac{1}{2}^{-}$ states
by enhancing combined  $^{3}\emph{PF}_{2}$ and 
$^{3}\emph{P}_{1}$ waves. However such a binding 
is a consequence of strongly resonant dineutrons in both mentioned waves. These
resonances are very sensible to the small reductions of the enhancement factor and  
they quickly vanish leaving only the dineutron ones.

\bigskip
In order to systematically explore the possible existence of  resonances
in all the three-neutron states, we let unchanged the original \emph{NN}-interaction and
force the binding by means of a phenomenological attractive
three-body force.
We have assumed the latter to have an hyperradial Yukawa form:
\begin{equation} \label{Yukawa_3NF}
V_{3n}=-W\frac{e^{-{\rho\over\rho_0}}}{\rho},\text{ with }\rho=\sqrt{%
x_{ij}^{2}+y_{ij}^{2}}  
\end{equation}
with $\rho_{0}=2$ fm. In this way, dineutron physics is not affected.

\begin{table}[h!]
\caption{Critical strengths $W_0$ in MeV$\cdot $fm of the phenomenological
Yukawa-type force of eq. (\ref{Yukawa_3NF}) required
to bind three-neutron in various states. Parameter $\protect\rho_{0}$
of this force was fixed to 2 fm. $W' $  are the values at which
three-neutron resonances become subthreshold ones, whereas $B_{trit}$ are
such 3NF corresponding triton binding energies in MeV.}
\label{Tab:Crit_val_w0}%
\begin{tabular}{l|llllll}
\hline
$J^{\pi}$ & $\frac{1}{2}^{+}$ & $\frac{3}{2}^{+}$ & $\frac{5}{2}^{+}$ & $%
\frac{1}{2}^{-}$ & $\frac{3}{2}^{-}$ & $\frac{5}{2}^{-}$ \\ \hline\hline
$W_0 $      & 307   & 1062 & 809 & 515 & 413 & 629 \\
$W' $       & 152   &    - & 329 & 118 & 146 & 277 \\
$B_{trit} $ & 21.35 &    - &  44.55 & 17.72 & 20.69& 37.05 \\\hline
\end{tabular}%
\end{table}

\begin{figure}[h!]
\begin{center}
\includegraphics[width=14.cm]{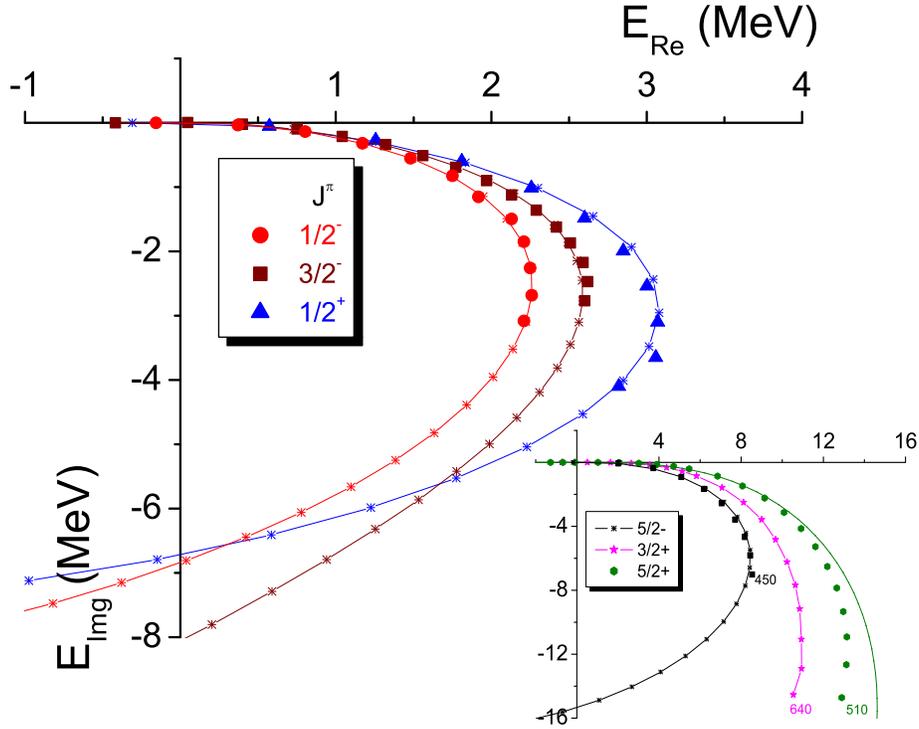} \vspace{-1.2 cm}
\end{center}
\caption{Three-neutron resonance trajectories obtained when varying the
strength $W$ of  Yukawa-type 3NF. Results
obtained using CS method are plotted with solid points.
For the $J^\protect\pi=1/2^-$ state,  $W$ varies from 520 
to 300 MeV$\cdot $fm by steps of 20 MeV$\cdot $fm, denoted for shortness [520,300,20]. 
For $J^\protect\pi=3/2^-$,
 W is in the range $[420,270,10]$ and for  $J^\protect\pi=1/2^+$  
in [310,210,10]. 
Other 3n states -- depicted in the smaller figure  -- require considerably stronger
3NF to be bound and result in large resonance energies values. 
For $J^\protect\pi=5/2^-$ 
$W$ changed in [610,450,10], for $J^\protect\pi=5/2^+$ $W$ in
[810,750,10] and then in [750,510,20], and for $J^\protect\pi=3/2^+$
in [1060,800,20] and [800,640,40].
ACCC method results are presented by solid lines, supported by star points.}
\label{Fig:3n_res_traj}
\end{figure}

In table~\ref{Tab:Crit_val_w0} we summarize the critical values $W_{0}$ of
the strength parameter $W$ for which the three-neutron system
is bound in different states. 
Corresponding resonance trajectories obtained by gradually
reducing parameter $W$ are traced in
Fig.~\ref{Fig:3n_res_traj}. As in previous plots, CS results are presented
by separate solid points, whereas ACCC ones are using continuous line
and star points.
One can see a very nice agreement between the two methods except for the
$J^{\pi}=\frac{5}{2}^{+}$ state, where a discrepancy between
two methods manifests at large energy. 
This is probably an artifact of the very strong 3NF used. 
Such a force confines the three-neutron system inside a $\approx$1.4 fm box, 
a distance smaller than the 3NF range itself, and starts to compete with
the hard-core repulsive part of the \emph{nn} interaction, making the ACCC method
badly convergent for broad resonances.
For the $J^{\pi}=\frac{3}{2}^{+}$
state, due to the even larger $W$ values, ACCC method has not been used.

One can remark that resonance trajectories have similar shapes for all
the three-neutron states, with the resonance poles tending to slip into
the third  quadrant ($\varepsilon_{res}<0$) well before $W$ equals 0, i.e.
when the additional 3NF are fully removed and only \emph{NN}-interaction remains.
In Table~\ref{Tab:Crit_val_w0} are also given the values
$W'$, obtained using ACCC method, at which ${^3n}$
resonances become subthreshold ($\varepsilon_{res}=0$). These values
are still rather large, strongly exceeding what could be
expected for realistic 3NF. To illustrate how strongly such 3NF
violate the nuclear properties, we give in the last row of the Table,
the triton binding energies,
obtained supposing that the same 3NF with strength $W'$ acts on it. 
These energies would be even larger if the range parameter $\rho_0$ 
taken in our 3NF model had had smaller and more realistic values.

The preceding results demonstrate that realistic \emph{NN}-interaction models exclude the
existence of observable three-neutron resonances. 
In~\cite{Csoto} ${^3n}$, resonance in $\frac{3}{2}^{+}$ state was claimed 
at $E=14-13i$ MeV for
the non-realistic Minnesota potential. Our results using realistic \emph{nn}
interaction contradict  however the existence of such a resonance. 
A very strong additional interaction is required to bind three-neutron in
$J^\pi=\frac{3}{2}^{+}$ state. Removing this interaction, the
imaginary part of the resonance grows very rapidly while
its real part saturates rather early (starting with $\approx$1060 MeV$\cdot $fm,
it reaches its maximal value at $W\approx$720 MeV$\cdot $fm). 
Once this saturation point is reached,
the resonance trajectory moves rapidly into the third quadrant.

\bigskip
The results shown in Figure~\ref{Fig:3n_res_traj} represent the
$^3n$ resonance trajectories only partially, 
without following them
to their final positions $W=0$, i.e. when the additional interaction is completely removed. 
The reason
is that these positions are very far from the bound state region, thus
requiring many terms in Pad\'e expansion to ensure an accurate ACCC result.
In principle, one could always imagine that these trajectories turn 
around and return to the fourth quadrant with positive real parts.
Although we have never encountered such a scenario in practical calculations,
we are not aware of any rigorous mathematical proof forbiding it
and it cannot consequently be excluded.

Such a kind of evolution seems however very unlikely in a physical case of interest.
The resonance trajectories
should indeed exhibit a very sharp behavior once entered in the third quadrant
whereas all our results indicate always a rather smooth variation.

\bigskip
In this study we have deliberately omitted the use of  realistic
3NF models. The reason is that such forces are not completely
settled yet, specially for pure neutron matter. 
In addition, we should remark that the
UIX~\cite{UIX} 3NF acts repulsively for multineutron
systems~\cite{Thesis}. The more recent Illinois 3NF contains charge
symmetry breaking (CSB) and considerably improves the underbinding
problem of neutron rich nuclei present for AV18+UIX~\cite{PPWC_PRC64}.
However, even strongly CSB realistic 3NF  would be by
order too weak to make three-neutron system resonant.
The reason of weak 3NF efficiency in multineutron physics is that such force requires
configurations when all three neutrons are close to each other, whereas such
structures are strongly suppressed by Pauli principle.

\begin{figure}[h!]
\begin{center}
\includegraphics[width=14.cm]{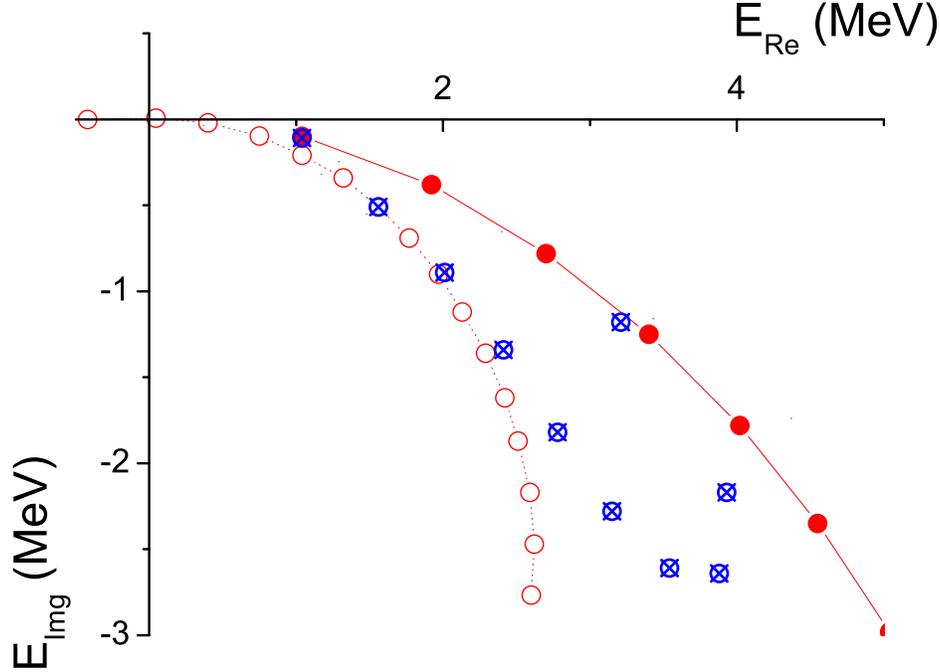} \vspace{-1.2 cm}
\end{center}
\caption{$J^\protect\pi=3/2^-$ 3n resonance trajectories
obtained when reducing the strength $W$ in the 
Yukawa-type 3NF (open circles) and enhancing $
^{3}\emph{PF}_{2}$ \emph{nn} interaction (full circles). Crossed
circles indicate the resonance path obtained when starting from 
$(W=360 \mbox{ MeV$\cdot$fm},\gamma=1)$ and going gradually to $(W=0,\gamma=3.7)$.}\label{Fig:3_2_nnn_identity}
\end{figure}

\bigskip
Finally, one can expect that the different ways
used to  artificially generate bound states 
could give rise to different resonance trajectories. 
Some resonance could thus have existed which are missed in our
approach. To investigate such a possibility we have chosen a  
$J^{\pi}=\frac{3}{2}^{-}$ resonance, obtained by means of the phenomenological 3NF
force (\ref{Yukawa_3NF}) with $W=360$ MeV$\cdot $fm. 
Then we gradually reduce $W$ to zero,  increasing at the same time 
the enhancement factor for the $^{3}\emph{PF}_{2}$
potential from $\gamma=1$ to $\gamma=3.7$. 
The resonance trajectory obtained this way is plotted (cross circles)
in Figure~\ref{Fig:3_2_nnn_identity}   together with
the resonance curves obtained by additional  3NF only (open circles) and by
enhancing only $^3\emph{PF}_2$ \emph{nn} 
potential (full circles). 
Once 3NF was completely removed, the resonance pole
joined the curve obtained with enhanced $^3\emph{PF}_2$ channel. 
Note that the structure of bound state obtained with 3N force
and by enhancing  $\emph{P}$-waves are quite different. 3NF requires very
dense and spherical symmetric neutron wave functions. This is the reason why
the $\frac{1}{2}^{+}$ state is more favorable than $\frac{3}{2}^{-}$ (see
Table~\ref{Tab:Crit_val_w0}) to bind three-neutron when using such an additional force.

\section{Conclusion}\label{sec:Conclu}

A systematic study of three-neutron resonances using realistic \emph{NN}-interaction
models was presented. The search of resonance positions was carried out
by artificially enhancing the interaction between the neutrons in such a
way that the three-neutron system becomes bound. 
Then, by gradually removing the additional
interaction we followed the path of the resonance energy located in the third and fourth
quadrants of the second Riemann sheet. 

Two different methods
were successfully applied to trace resonance trajectories, 
namely the Complex Scaling (CS) and the Analytical Continuation
in the Coupling Constant (ACCC). They provided results in very good agreement. 
Two alternative
ways were also explored to enhance the interaction between neutrons: 
in one of them, the three-neutron system 
was bound by adding a phenomenological three-nucleon force (3NF)
and in the other one the binding was obtained  by enhancing the interaction in
some \emph{NN} partial waves. 

All ${^3n}$ resonance trajectories, for  states
up to $J=5/2$,  were shown to move into the third energy quadrant (Re(E)$<0$) 
becoming subthreshold resonances,
well before the additional interaction is fully removed.
Our results clearly demonstrate that the current realistic interactions 
exclude the possible existence
of bound as well as resonant three-neutron states. 
To push the three-neutron resonance out of the subthreshold region, 
such models should already be strongly violated. 
These findings support the results of  Ref~\cite{Glockle_Hem}
for simplified \emph{NN}-interaction model.

The possible existence of observable
four-neutron (tetraneutron) resonance seems rather doubtful as well, 
since for such a system to be artificially bound one requires almost as 
large enhancement factors as
in the three-neutron case~\cite{Thesis}. 
Still, such a possibility can not be
completely excluded, since the tetraneutron is favored due to presence of two
almost bound dineutron pairs. 
Moreover, recent experiment on $^8He(d,{^6Li})4n$
reaction shows an excess of low energy $^6$Li nuclei with respect to 
what one could expect from a phase space analysis~\cite{Exxon}. 
The presence of events associated
with a resonant tetraneutron state was suggested.
The possible existence of such structures will be explored in a forthcoming work.

\bigskip
{\bf Acknowledgements:}
Numerical calculations were performed at
Institut du D\'eveloppement et des Ressources en Informatique Scientifique (IDRIS) from  CNRS
and at Centre de Calcul Recherche et Technologie (CCRT) from CEA Bruy\`eres le Ch\^atel.
We are grateful to the staff members of these two organizations for their kind hospitality and useful advices.

\end{document}